%% file: cond.tex
\documentclass[11pt]{article}
\usepackage{epsfig,sint,macros,cite,mathbbol}

\begin{document}
\input title.tex
\input sect1.tex

\input sect2.tex

\input sect3.tex

\input sect4.tex
\input sect5.tex

\input sect6.tex

\input sect7.tex


\input biblio.tex
\end{document}

%% file: title.tex
\begin{titlepage}
\begin{flushright}
DESY 05-129
\end{flushright}

\vskip 0.5 cm
\begin{center}
  {\Large\bf 
  On the renormalized scalar density in quenched QCD \\[0.5ex] }
\end{center}
\vskip 0.5 cm
\begin{center}
{\large 
     J. Wennekers and H. Wittig
}
\vskip 0.5cm
DESY, Theory Group, Notkestrasse 85,\\
D-22603 Hamburg, Germany
\vskip 1.0cm
{\bf Abstract}
\vskip 0.35ex
\end{center}

\noindent
We present a non-perturbative determination of the renormalization
factor $\zs$ of the scalar density in quenched QCD with overlap
fermions. Results are obtained at four values of the lattice spacing.
By combining $\zs$ with results for the low-energy constant $\Sigma$
we are able to compute the renormalization group invariant scalar
condensate $\widehat\Sigma$ in the continuum limit with a total
accuracy of 7\%, excluding dynamical quark effects. Our result
translates to $\Sigma_\msbar(2\,\GeV)=(285\pm9\,\MeV)^3$ if the scale
is set by the kaon decay constant.  We have also performed scaling
studies of the pseudoscalar decay constant and the vector mass. Our
results indicate that quantities computed using overlap quarks exhibit
excellent scaling behaviour, with small residual lattice artifacts.

\vfill
\begin{center}
July 2005
\end{center}

\eject

\vfill

\eject

\end{titlepage}

%% file: sect1.tex
\section{Introduction}

The expectation value of the scalar density at vanishing quark mass,
commonly named the quark or chiral condensate, plays a central r\^ole
in QCD at low energies.  Spontaneous chiral symmetry breaking is
signalled by the formation of a non-vanishing condensate, and an
accurate determination of its value is of great practical interest.
Lattice simulations of QCD appear well suited for this task, but in
order to guarantee a reliable error estimation, it is crucial to have
control over systematic effects. In particular, to ensure that the
quark condensate approaches the continuum limit as a power series in
the lattice spacing~$a$, the renormalization of the bare scalar
density must be known with good accuracy. It is well known that
renormalization factors computed in perturbation theory at one loop
are not reliable.  Further complications arise if the lattice
formulation breaks chiral symmetry explicitly. For instance, in the
case of Wilson fermions, a cubically divergent term must be subtracted
before multiplicative renormalization can be applied
\cite{chiral_latt}.

In this paper we report on a non-perturbative calculation of the
renormalization factor $\zs$ of the scalar density, using the overlap
operator \cite{NeubergerDirac} as our fermionic discretization in the
quenched approximation. We employ the method proposed in \cite{HJLW}
and compute $\zs$ at four different values of the lattice spacing,
ranging from $a\approx0.12\,\fm$ to $0.075\,\fm$. By identifying the
bare condensate with the low-energy constant $\Sigma$, which appears
in effective low-energy descriptions of QCD, we can compute the
renormalized quantity, given results for $\Sigma$ at the corresponding
values of the bare coupling in the quenched theory.

Our analysis of the scaling properties of the renormalized condensate
indicates the presence of only very small cutoff effects of
order~$a^2$, provided that the non-perturbative estimates for the
renormalization factor $\zs$ are used throughout. Thus, an
extrapolation to the continuum limit can be performed in a controlled
way.

Moreover, we have extended the scaling analysis to other quantities,
such as the pseudoscalar meson decay constant and the mass in the
vector channel. In all cases we observe an excellent scaling
behaviour, with leading cutoff effects of order $a^2$, and thus
consistent with expectation. To our knowledge, these results represent
the first detailed scaling study for overlap fermions.

Results for the quark condensate have already been published by a
number of authors
\cite{APE_cond,JLW_cond,RBC_cond,DeG_cond,GHR_cond,HJLW_lat01,Bern_cond,
  BeciLub_cond,GLMPR_cond,McNeile_cond}.  The novelty in this paper is
the extension of previous simulations with overlap fermions
\cite{JLW_cond,DeG_cond,GHR_cond} to considerably finer lattice
spacings, as well as the strict application of non-perturbative
renormalization, enabling us to take the continuum limit. Overlap
fermions, despite their larger numerical cost, have clear conceptual
advantages when it comes to studying the problem of chiral symmetry
breaking, which is encoded in the value for the quark condensate. We
stress, though, that our results are valid for quenched QCD, and thus
great care must be taken if they are to be interpreted in the context
of the full theory. In particular, the  chiral condensate is
ill-defined in the quenched approximation \cite{Quen_Chiral}.

%% file: sect2.tex
\section{Renormalization conditions}

Here we briefly recall the conditions that fix the renormalization of
the scalar and pseudoscalar densities in simulations using fermionic
discretizations that satisfy the Ginsparg-Wilson relation
\cite{GinsWil,ExactChSy}.  Full details can be found in
refs.\,\cite{HJLW,HJLW_lat01}.

If the regularization preserves chiral symmetry, then the chiral Ward
identities imply that
\be
    \zs=\zp=1/\zm.
\ee
The renormalization factor $\zshat$, which relates the bare scalar
density to the renormalization group invariant (RGI) density, can then
be defined by \cite{HJLW}
\be
   \zshat(g_0) = \left.
   \frac{(r_0\,m)(g_0)}{\UM}\right|_{(r_0\,\mps)^2=\xref}. 
\label{eq_zsUM_def}
\ee
In this expression $\UM$ denotes the RGI quark mass in the continuum
limit, in units of the hadronic radius $r_0$ \cite{r0_refs}, while $m$
is the bare quark mass that appears in the lattice Dirac operator
satisfying the Ginsparg-Wilson relation. The expression on the
right is evaluated at a given reference value, $\xref$, of the square
of the pseudoscalar meson mass in units of $r_0$. A convenient choice,
which we also adopt here, is $\xref=1.5736$. For $r_0=0.5\,\fm$ this
corresponds to $\mps=\mk=495\,\MeV$. The original data required for
the determination of $\UM$ were published in \cite{mbar:pap3}, and in
eq.~(3.1) of \cite{HJLW} $\UM$ is listed for several choices of
$\xref$. 

Since $\zs=\zp$ an alternative renormalization condition can be
formulated in terms of the matrix element of the pseudoscalar
density. If we introduce the shorthand notation
\be
   \Gpb=\langle0|P^a(0)|{\rm{PS}}\rangle,\qquad
   P^a(x)=(\psibar\lambda^a\gamma_5\psi)(x), 
\ee
where $\lambda^a$ is some flavour matrix, then $\zphat$ can be defined
via
\be
   \zphat = \left.
   \frac{\UP}{(r_0^2\Gpb)(g_0)}\right|_{(r_0\,\mps)^2=\xref}.
\label{eq_zsUP_def}
\ee
The universal factor $\UP$ denotes the RGI matrix element of the
pseudoscalar density in the continuum limit. Its value can be
determined, for instance, using $\rmO(a)$ improved Wilson fermions,
and the results presented in refs.\,\cite{mbar:pap1,mbar:pap3} then
yield
\be
   \UP = 1.802(42) \qquad\hbox{at}\quad (r_0\,\mps)^2=1.5736.
\ee
In order to compute $\zshat$ or $\zphat$, it is clear from
eqs.\,(\ref{eq_zsUM_def}) and\,(\ref{eq_zsUP_def}) that
the main task is the determination of the value of the bare quark
mass, $m$, and the matrix element $\Gpb$ at the point where
$(r_0\,\mps)^2=\xref$, for a fermionic discretization based on the
overlap operator.

%% file: sect3.tex
\section{Numerical simulations}

In our simulations we have computed mesonic two-point correlation
functions in the pseudoscalar and vector channels. We have used the
massive overlap operator $D_m$, defined by \cite{NeubergerDirac}
\be
   D_m=\left(1-\half\abar{m}\right)D+m,\qquad
   D=\frac{1}{\abar}\left(1-\frac{A}{\sqrt{A^\dagger{A}}}\right),
\label{eq_Ddef}
\ee
where
\be
   A=1+s-aD_{\rm w},\qquad \abar=\frac{a}{1+s},\qquad |s|<1,
\label{eq_Adef}
\ee
and $D_{\rm w}$ is the Wilson-Dirac operator.

The calculation of the quark propagator proceeds as usual by solving
\be
   D_m\psi = \eta
\label{eq_Dpsi_eta}
\ee
for a source field $\eta$. As pointed out in \cite{numeps}, the
determination of both chiralities of the solution $\psi$ requires some
care in the presence of zero modes of the massless operator $D$,
especially as the quark mass becomes small. To separate off the zero
mode contribution we have implemented the strategy outlined in
section\,7 of ref.~\cite{numeps}, which we briefly review here. To this
end we shall consider a gauge configuration which has a number of zero
modes with positive chirality. 

The solution to \eq{eq_Dpsi_eta} with negative chirality is given by
\be
   P_{-}\psi = (D_m^\dagger D_m)^{-1} P_{-}D_m^\dagger\eta,
\label{eq_Pm_psi}
\ee
and thus the inversion of $D_m^\dagger D_m$ takes place in the
chirality sector that does not contain zero modes. The components with
positive chirality are obtained from
\be
   P_{+}\psi=\frac{1}{m}P_0P_{+}\eta + (P_{+}D_m
   P_{+})^{-1}\Big\{(1-P_0)P_{+}\eta -P_{-}D_m P_{-}\psi\Big\},
\label{eq_Pp_psi}
\ee
where $P_0$ is a projector onto the subspace spanned by the zero
modes, and whose calculation is described in \cite{numeps}. When
implemented in a computer program, \eq{eq_Pp_psi} offers complete
control over the zero mode contribution. It is also clear that the
necessary inversion of $(P_{+}D_m P_{+})$ is performed on a source
where all zero mode contributions have been projected out. The r\^oles
of the positive and negative chirality sectors are obviously reversed
in the above expressions if the zero modes have negative chirality.

In our programs we compute $P_{-}\psi$ and $P_{+}\psi$ using the
Generalized Minimum Residual (GMRES) algorithm \cite{YSaad}, which
also allows for an inversion of $D_m$ itself.  To speed up the
inversion we have incorporated ``low-mode preconditioning'', a
technique designed to protect against numerical instabilities caused
by very small eigenvalues of $D_m^\dagger D_m$ \cite{numeps}. As we shall
see later, the quark masses considered in this work are relatively
large and hence provide an infrared cutoff, but we found that the
inversion can nevertheless be accelerated in this way. The presence of
zero modes in conjunction with the fact that the low (non-zero) modes
are only known with a certain numerical accuracy requires some care in
the implementation of low-mode preconditioning for the solution in
eq.~(\ref{eq_Pp_psi}). Details will be described elsewhere
\cite{JW_thesis}.

Since the goal of our study is the computation of the renormalized
condensate in the continuum limit, we have chosen our simulation
parameters to coincide with those of previous determinations of the
bare condensate. To this end we have identified the latter with the
parameter $\Sigma$ computed by matching the spectrum of low-lying
eigenvalues of $D$ to the predictions of Random Matrix Theory
\cite{rmt}. More precisely, we have concentrated on the dataset
labeled ``B'' in that reference, which comprises three different
lattice spacings at a fixed box size of $L=1.49\,\fm$. We note that a
spatial volume of this size is sufficiently large to avoid large
finite volume effects for masses and decay constants at
$\mps\approx\mk$. In order to improve the accuracy of the continuum
extrapolation we added a fourth $\beta$-value, $\beta=5.9256$, tuned
to reproduce the same physical box size for $L/a=14$. Following the
same procedure as in \cite{rmt}, we have determined the low-lying
spectrum of the Dirac operator and extracted the parameter $\Sigma$.

The computation of fermionic two-point functions proceeded by setting
$T=2L$, to control the exponential decay of the correlation function
in a more reliable way. Our simulation parameters are listed in
Table~\ref{tab_simpar}. As in ref. \cite{rmt}, the parameter~$s$ in
the definition of the overlap operator (c.f. \eq{eq_Adef}) was set to
$s=0.4$.
\begin{table}[ht]
  \begin{center}
    \vspace{0.25cm}
    \begin{tabular}{ccccc}
      \hline \\[-2.0ex]
      $\beta$ & $L/a$ & $r_0/a$ & $a\;[\fm]$ & $\#$cfgs \\[0.7ex]
      \hline \\[-2.0ex]
      $5.8458$ & $12$ & $4.026$ & 0.124 & 200 \\
      $5.9256$ & $14$ & $4.697$ & 0.106 & 174 \\
      $6.0000$ & $16$ & $5.368$ & 0.093 & 200 \\
      $6.1366$ & $20$ & $6.710$ & 0.075 & 100 \\
      \hline \\[-2.0ex]
    \end{tabular}
  \end{center}
  \caption{\footnotesize Simulation parameters for the determination
           of mesonic two-point functions \label{tab_simpar}}
\end{table}

At each value of the bare coupling we computed quark propagators for
three bare masses straddling the reference point corresponding to
$\xref=(r_0\mps)^2=1.5736$. We added a fourth, heavier value at all
but the finest lattice spacing we considered, to study the quark mass
dependence of mesonic quantities in more detail. Since the quark
masses here are relatively large, the low-lying spectrum of $D$ cannot
induce large fluctuations in correlation functions like those observed
in the so-called $\epsilon$-regime \cite{HUB_eps,lma}. Therefore, we did
not apply the method known as low-mode averaging \cite{lma,DeG_Schaef}
to enhance the signal.

In the pseudoscalar channel we used both the left-handed axial current
$J_\mu$ and the pseudoscalar density $P$ as interpolating operators,
i.e.
\be
   J_\mu(x) = (\psibar_r\gamma_\mu P_{-}\psi_s)(x),\qquad
   P(x) = (\psibar_r\gamma_5\psi_s)(x),
\ee
where $P_\pm=\half(1\pm\gamma_5)$, and $r,\,s$ denote flavour
labels. Choosing $r\neq{s}$, both of these composite fields were then
combined into non-singlet two-point correlation functions
\be
   C_{\rm QR}(x_0) = a^3\sum_{\xvec}\left\langle Q(x) R(0)
   \right\rangle, \qquad Q,\,R = J_0,\,P.
\ee
The correlation function $C_{\rm JJ}$ involves only the left-handed
quark propagator such that zero modes cannot contribute. By contrast,
$C_{\rm PP}$ includes components of the quark propagator whose
chirality coincides with that of the zero modes (if any). The latter
can be separated off by implementing the expression in \eq{eq_Pp_psi}.

The pseudoscalar mass and decay constant, as well as the matrix
element $\Gpb$ were extracted from single-cosh fits, after averaging
the correlators over the forward and backward halves of the
lattice. Good plateaus were observed, which served as a guideline for
choosing our fit intervals. We also computed the current quark mass,
$\mpcac$ from
\be
   a\mpcac=\frac{1}{2}
   \frac{\half(\partial_0+\partial_0^*)C_{\rm JP}(x_0)}{C_{\rm
   PP}(x_0)},
\ee
where $\partial_0,\,\partial_0^*$ denote the forward and backward
lattice derivatives, respectively. In order to compute meson masses
in the vector channel, we have considered the two-point correlator
\be
   C_{\rm VV}(x_0) = a^3\sum_{\xvec}\sum_{k=1}^3 \left\langle
   V_k(x)V_k(0)\right\rangle,\qquad V_k(x) =
   (\psibar_r\gamma_k\psi_s)(x),\qquad k=1,2,3. 
\ee
It turned out to be impossible, however, to obtain a stable plateau
for the effective mass, by simply using a local source vector in the
inversion step. Therefore we applied Jacobi smearing on the source
$\eta$, as described in \cite{jacobi}. The parameters were chosen such
that the rms.~smearing radius in units of $r_0$ was kept constant at
approximately 0.6. With this choice we were able to improve the
stability of the plateau in the vector channel considerably.

%% file: sect4.tex
\section{Determination of renormalization factors}

Our results for masses and matrix elements are summarized in Table
\ref{res_tab}.
\begin{table}[ht]
  \begin{center}
    \vspace{0.25cm}
    \begin{tabular}{ccccccc}
      \hline \\[-2.0ex]
      $\beta$ & $am$ & $a\mpcac$ & $a\mps$ & $a\Fpb$ &
      $a^2\Gpb$ & $a\mv$ \\[0.7ex]
      \hline \\[-2.0ex]
      $5.8458$ & $0.040$ & $0.02359(6)$ & $0.262(9)$ & $0.0417(10)$
               & $0.1185(61)$ & $0.532(37)$ \\
               & $0.053$ & $0.03134(7)$ & $0.294(8)$ & $0.0424(9)$
               & $0.0889(39)$ & $0.537(31)$ \\
               & $0.067$ & $0.03973(8)$ & $0.327(8)$ & $0.0434(8)$
               & $0.0718(28)$ & $0.556(24)$ \\
               & $0.113$ & $0.06769(11)$ & $0.421(6)$ & $0.0469(7)$
               & $0.0488(14)$ & $0.631(14)$ \\[0.7ex]
      $5.9256$ & $0.034$ & $0.02120(13)$ & $0.235(7)$ & $0.0389(10)$
               & $0.0877(39)$ & $0.502(21)$\\
               & $0.046$ & $0.02875(14)$ & $0.266(6)$ & $0.0397(10)$
               & $0.0657(26)$ & $0.515(15)$ \\
               & $0.057$ & $0.03569(15)$ & $0.292(6)$ & $0.0405(9)$
               & $0.0547(19)$ & $0.529(12)$ \\
               & $0.097$ & $0.06120(17)$ & $0.377(4)$ & $0.0433(9)$
               & $0.0378(11)$ & $0.579(7)$ \\[0.7ex]
      $6.0000$ & $0.030$ & $0.01927(7)$ & $0.217(6)$ & $0.0346(7)$
               & $0.0814(42)$ & $0.424(15)$ \\
               & $0.040$ & $0.02576(7)$ & $0.247(5)$ & $0.0356(6)$
               & $0.0612(27)$ & $0.445(11)$ \\
               & $0.050$ & $0.03229(7)$ & $0.273(5)$ & $0.0366(6)$
               & $0.0501(20)$ & $0.462(9)$ \\
               & $0.085$ & $0.05543(9)$ & $0.352(3)$ & $0.0403(5)$
               & $0.0342(10)$ & \\[0.7ex]
      $6.1366$ & $0.024$ & $0.01638(6)$ & $0.168(5)$ & $0.0296(7)$
               & $0.0447(21)$ & $0.360(28)$ \\
               & $0.032$ & $0.02185(6)$ & $0.195(4)$ & $0.0301(6)$
               & $0.0356(14)$ & $0.378(20)$ \\
               & $0.040$ & $0.02734(6)$ & $0.218(4)$ & $0.0309(6)$
               & $0.0305(11)$ & $0.389(15)$ \\
      \hline \\[-2.0ex]
    \end{tabular}
  \end{center}
  \caption{\footnotesize Results for meson masses and decay constants
  computed at several values of quark masses at each lattice
  spacing. The results for $a\mps$ and $a\Fpb$ were extracted from
  correlators of the left-handed axial current. \label{res_tab}}
\end{table}

\begin{figure}[t]
    \centerline{\includegraphics*[width=8cm]{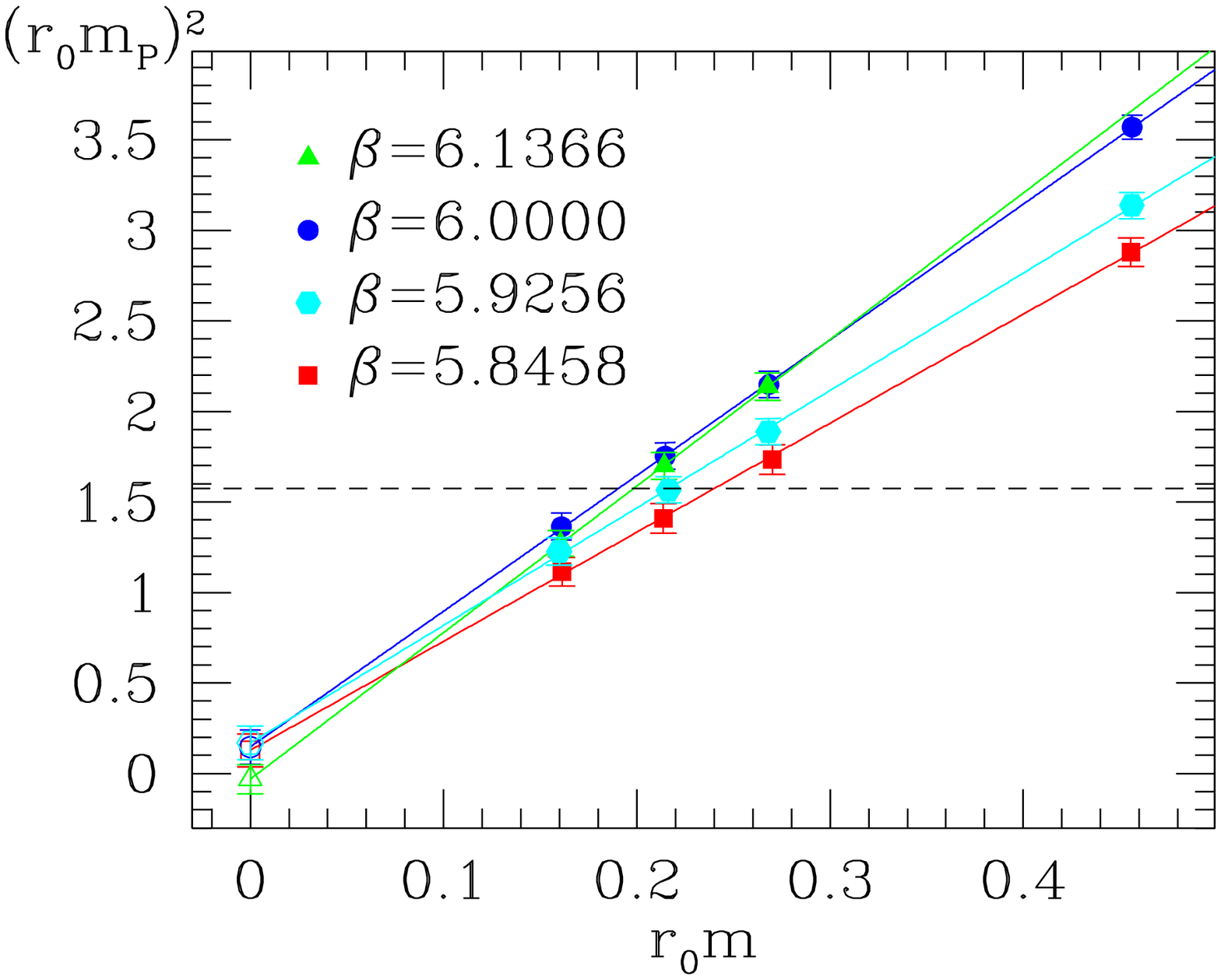}}
    \caption{\footnotesize $(r_0 \mps)^2$ as a function of $r_0
    m$. The horizontal dashed line represents the reference point
    $(r_0\mps)^2=1.5736$. \label{fig_mps2_m}}
\end{figure}

In order to compute $\zshat$ according to \eq{eq_zsUM_def} we have to
determine the quark mass at the reference point in units of $r_0$. In
Fig.~\ref{fig_mps2_m} we have plotted $(r_0\mps)^2$ as a function of
$(r_0m)$ at all four $\beta$-values. As can be seen, the data are
easily fitted by straight lines, but a non-zero intercept is found at
all but the largest value of $\beta$: the pseudoscalar mass at zero
bare quark mass differs from zero by $1-2$ standard deviations. Since
the correlation function of the left-handed axial current is free from
contributions of zero modes, they cannot be responsible for the
non-zero intercept. We note however that the chiral fits yield
$\chi^2/\rm dof$ below~1 even if the extrapolation is forced through
the origin.

By performing local interpolations to the reference point using the
three nearest data points and subsequently applying \eq{eq_zsUM_def},
we obtain the values of $\zshat$, which are tabulated at each
$\beta$-value in Table~\ref{tab_zs_res}. The typical accuracy of our
determination is around 5\%. It should be noted that the precision is
partly limited by the accuracy of the published value of $\UM$, which
is about 3\% \cite{mbar:pap3}. We estimate that pushing the precision
of our determination of $\zshat$ to that level would require a
four-fold increase in statistics.

\begin{table}
  \begin{center}
    \vspace{0.25cm}
    \begin{tabular}{cccc}
      \hline \\[-2.0ex] 
      $\beta$ & $\zshat$ & $\zphat$ & 
      $\za$\\[0.7ex]
      \hline \\[-2.0ex] 
      $5.8458$ & $1.28(6)$ & $1.33(4)$ & $1.710(5)$ \\
      $5.9256$ & $1.19(7)$ & $1.20(4)$ & $1.611(3)$ \\
      $6.0000$ & $1.05(5)$ & $0.88(6)$ & $1.553(2)$ \\
      $6.1366$ & $1.01(4)$ & $1.02(5)$ & $1.478(2)$ \\
      \hline \\[-2.0ex]
    \end{tabular}
  \end{center}
  \caption{\footnotesize Non-perturbative determinations of $\zshat$,
  $\zphat$ and $\za$.\label{tab_zs_res}}
\end{table}

\begin{figure}[ht]
  \begin{minipage}[t]{7cm}
    \centerline{\includegraphics*[width=7cm]{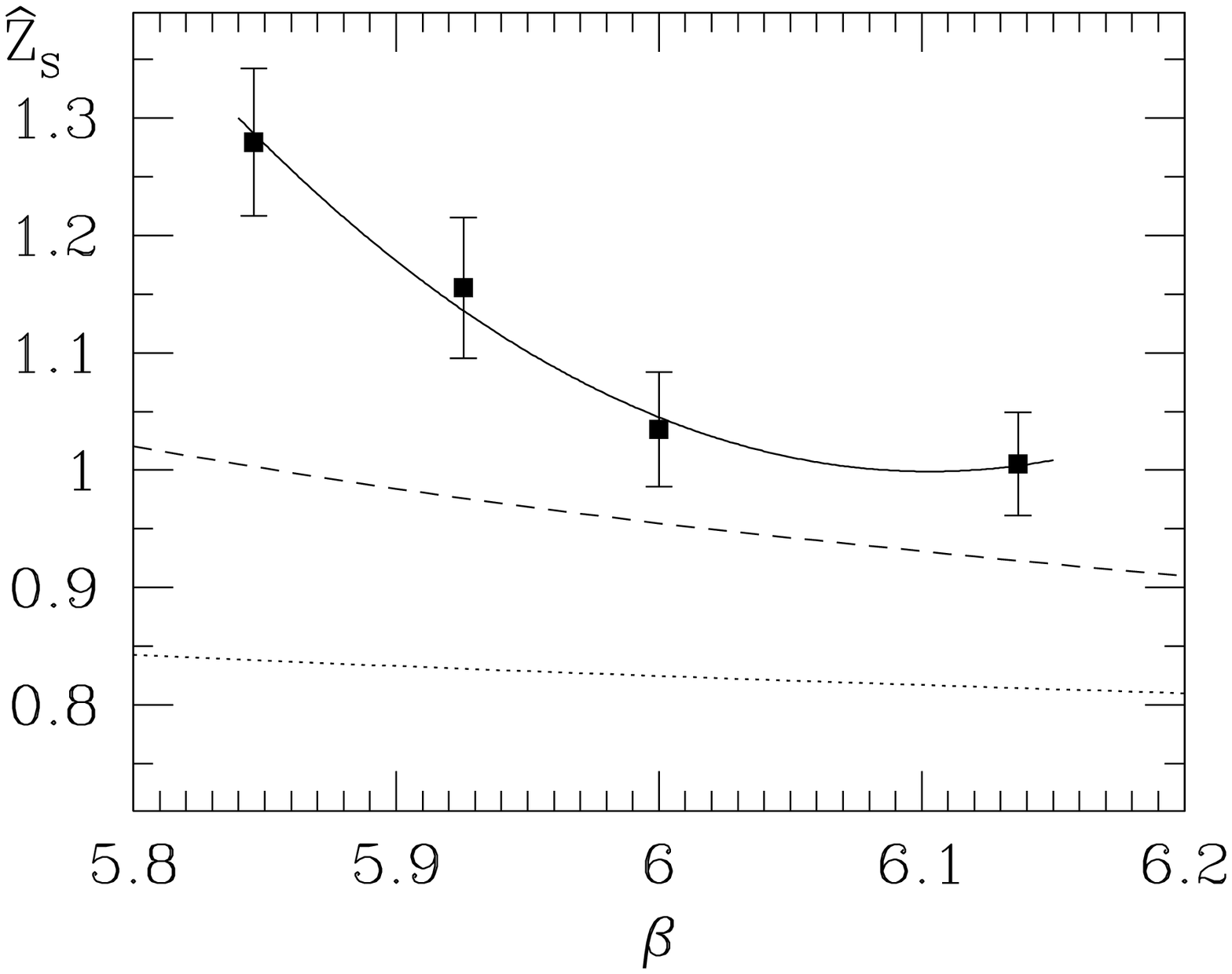}}
    \caption{\footnotesize $\zshat$ as a function of $\beta$. The
      solid line denotes the fit of \eq{zsfit}. The dotted and dashed
      curves represent the results of bare and mean-field improved
      perturbation theory at one loop order.\label{fig_zs_res}}
  \end{minipage}
  \hfill
  \begin{minipage}[t]{7cm}
    \centerline{\includegraphics*[width=7cm]{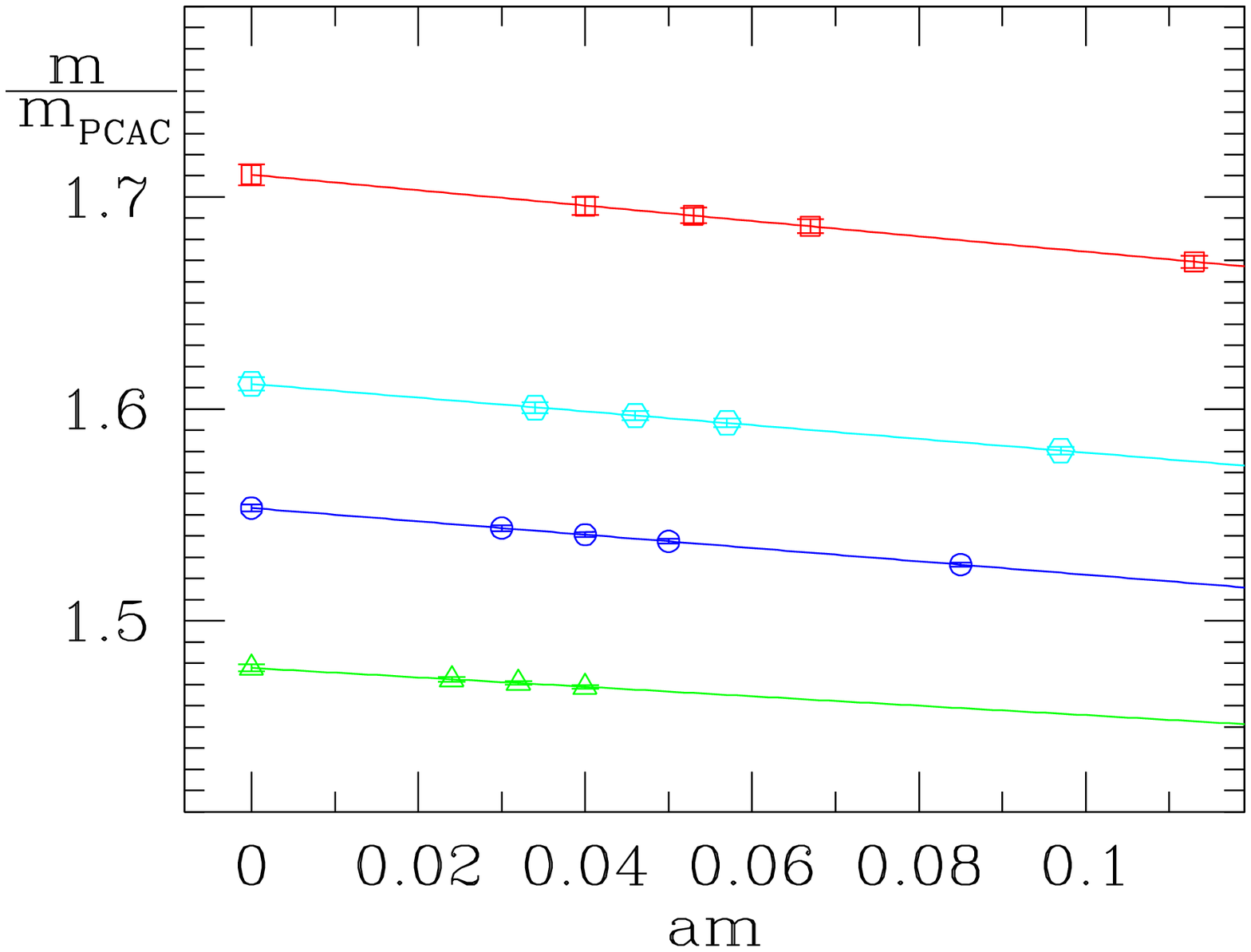}}
    \caption{\footnotesize The quark mass dependence of
             $\frac{m}{\mpcac}$. The value of $\beta$ increases from
             top to bottom. $\za$ is defined as the value of this
             ratio in the limit of vanishing quark
             mass.\label{fig_massrat}}
  \end{minipage}
\end{figure}

In Fig.~\ref{fig_zs_res} we plot our results for $\zshat$ versus
$\beta$. It has become customary to represent results for
renormalization factors at different values of the bare coupling by
interpolating curves. Using a simple polynomial ansatz in $(\beta-6)$
yields
\be
   \zshat(\beta)=1.045-0.899(\beta-6)+4.36(\beta-6)^2, \qquad s=0.4.
   \label{zsfit}
\ee
This formula describes $\zshat$ with an estimated error of 5\% in the
studied range of $\beta$, i.e. $5.8458\leq\beta\leq6.1366$. We
emphasize that our determination is valid only for the case $s=0.4$ in
the definition of the Neuberger-Dirac operator, eqs.~(\ref{eq_Ddef})
and~(\ref{eq_Adef}).

The perturbative expression for $\zshat$ at one loop is
\be
  \zshat^{\rm pt}(g_0) = \frac{\mbar_\msbar(\mu)}{M} \left\{
                 1+g_0^2\left[\frac{1}{2\pi^2}\ln(a\mu)
                 +z_{\rm S}^{(1)}\right]+\rmO(g_0^4)\right\},
\label{eq_zspbare}
\ee
where $z_{\rm S}^{(1)}=0.147107$ for our choice of $s=0.4$
\cite{chiral:AlFoPaVi,SteLeo00}. The factor $\mbar_\msbar(\mu)/M$ was
computed previously in \cite{mbar:pap3}. The mean-field improved
version of $\zshat^{\rm pt}$ reads \cite{HJLW}

\be
   \zshat^{\rm mf}(g_0) = \frac{\mbar_\msbar(\mu)}{M}
   \left(\frac{1+s}{1+\tilde s}\right)\left\{ 1+g^2
      \left[\frac{1}{2\pi^2}\ln(a\mu)+z_{\rm S}^{(1)}+u_0^{(1)}\left(
        \frac{3-s}{1+s}\right)
        \right]\right\}\ ,
\label{eq_zspttad}
\ee
where $g^2=g_0^2/u_0^4$ is the boosted coupling,
$\tilde{s}=3+(s-3)/u_0$, with $u_0^4$ being the average plaquette. The
comparison of our numerical results for $\zshat$ with perturbation
theory is shown in Fig.~\ref{fig_zs_res}. The mean-field improved
perturbative expansion comes quite close to the non-perturbatively
determined values for $\beta\;\gtaeq\;6.0$ but falls short by more
than 20\% below $\beta=6.0$. Unsurprisingly, perturbation theory in
the bare coupling $g_0^2$ fares a lot worse in the entire range of
couplings studied here.

The results for $\zphat$, computed according to \eq{eq_zsUP_def}, are
listed alongside those for $\zshat$ in Table~\ref{tab_zs_res}. The
renormalization conditions for $\zshat$ and $\zphat$ imply that the
two must be identical up to effects of order~$a^2$. Indeed, we observe
hardly any difference at our level of accuracy, except at
$\beta=6.0$. In our view, the most likely explanation for this
deviation is a statistical fluctuation.

In order to include the pseudoscalar decay constant in the scaling
tests described below we also computed the renormalization factor of
the axial current, $\za$. Using the PCAC relation and $\zm=1/\zp$ one
can define
\be
   \za = \lim_{m\to0}\frac{m}{\mpcac}. 
\ee
We found the ratio ${m}/{\mpcac}$ to depend only weakly on the bare
mass (c.f.~Fig.~\ref{fig_massrat}). $\za$ could then be determined by
extrapolating $m/\mpcac$ linearly in $m$ to the chiral limit.

%% file: sect5.tex
\section{The renormalized condensate}

Having determined the renormalization factor of the scalar density in
a range of bare couplings, we can now compute the renormalized
condensate in the continuum limit, by combining the results for
$\zshat$ with estimates of the bare condensate.

In effective low-energy descriptions of QCD with $\nf=3$ quark
flavours, the quark condensate is identified with the low-energy
constant $\Sigma$ via
\be
   -\left\langle\psibar\psi\right\rangle = \Sigma.
\ee
In the quenched theory, however, the condensate
$-\langle\psibar\psi\rangle$ is not defined, owing to the presence of
infrared divergencies as the chiral limit is approached
\cite{Quen_Chiral}. Nevertheless, the low-energy constant $\Sigma$ can
be determined in quenched QCD, for instance, by comparing lattice data
of suitable quantities to expressions of Chiral Perturbation Theory or
chiral Random Matrix Theory. Although in this case the identification
of $\Sigma$ with the quark condensate is rather dubious, we shall
nevertheless proceed to compute a renormalized ``condensate'', by
assuming that estimates of $\Sigma$ in the quenched theory renormalize
like the scalar density.

Our input quantities are thus the renormalization factors $\zshat$ of
Table~\ref{tab_zs_res} and results for $\Sigma$, determined by
matching the low-lying eigenvalues of the Dirac operator in the
$\epsilon$-regime to the predictions of the chiral unitary random
matrix model according to \cite{rmt}
\be
   \left\langle\lambda_k\right\rangle_{\nu}{\Sigma}V
  =\left\langle\zeta_k\right\rangle_{\nu},  \qquad k=1,2,\ldots
\label{eq_rmt_match}
\ee
Here, $\langle\lambda_k\rangle_{\nu}$ is the expectation value of the
$k$th eigenvalue in the topological sector with index $\nu$, and
$\zeta_k$ denotes the $k$th scaled eigenvalue in the matrix model. In
ref. \cite{rmt} it was found that good agreement with random matrix
behaviour is observed for lattice volumes $V$ of at least
$(1.5\,\fm)^4$. In other words, the value of $\Sigma$ extracted from
\eq{eq_rmt_match} depends neither on the particular eigenvalue, nor on
the topological sector, within statistical errors.

\begin{figure}[ht]
  \begin{minipage}[t]{7cm}
    \centerline{\includegraphics*[width=7cm]{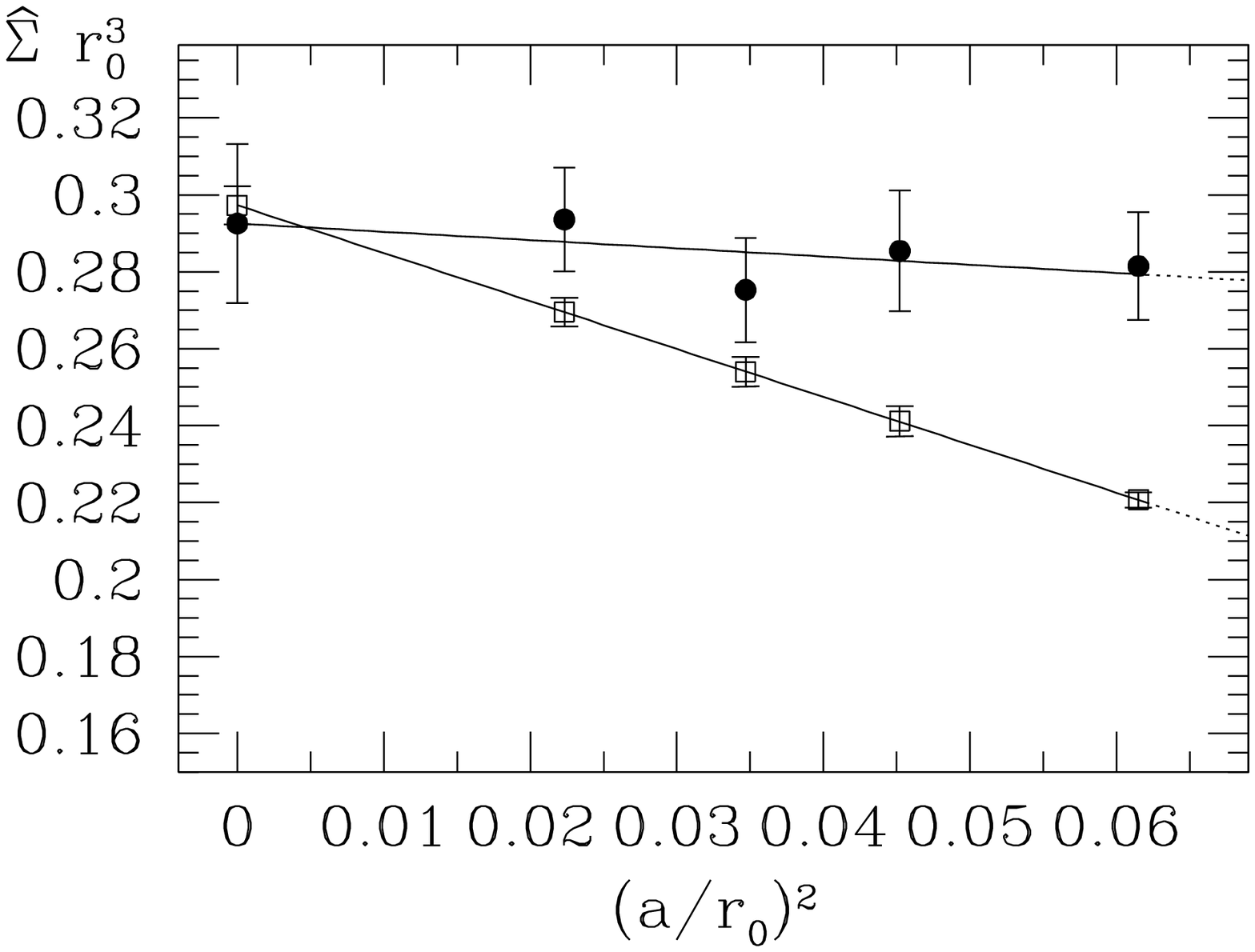}}
    \caption{\footnotesize Continuum extrapolation of
    $r_0^3\widehat\Sigma$. Full circles denote the results obtained
    using non-perturbative renormalization factors, while open squares
    represent values resulting from applying mean-field improved
    perturbation theory.\label{fig_zshat}}
  \end{minipage}
  \hfill
  \begin{minipage}[t]{7cm}
    \centerline{\includegraphics*[width=7cm]{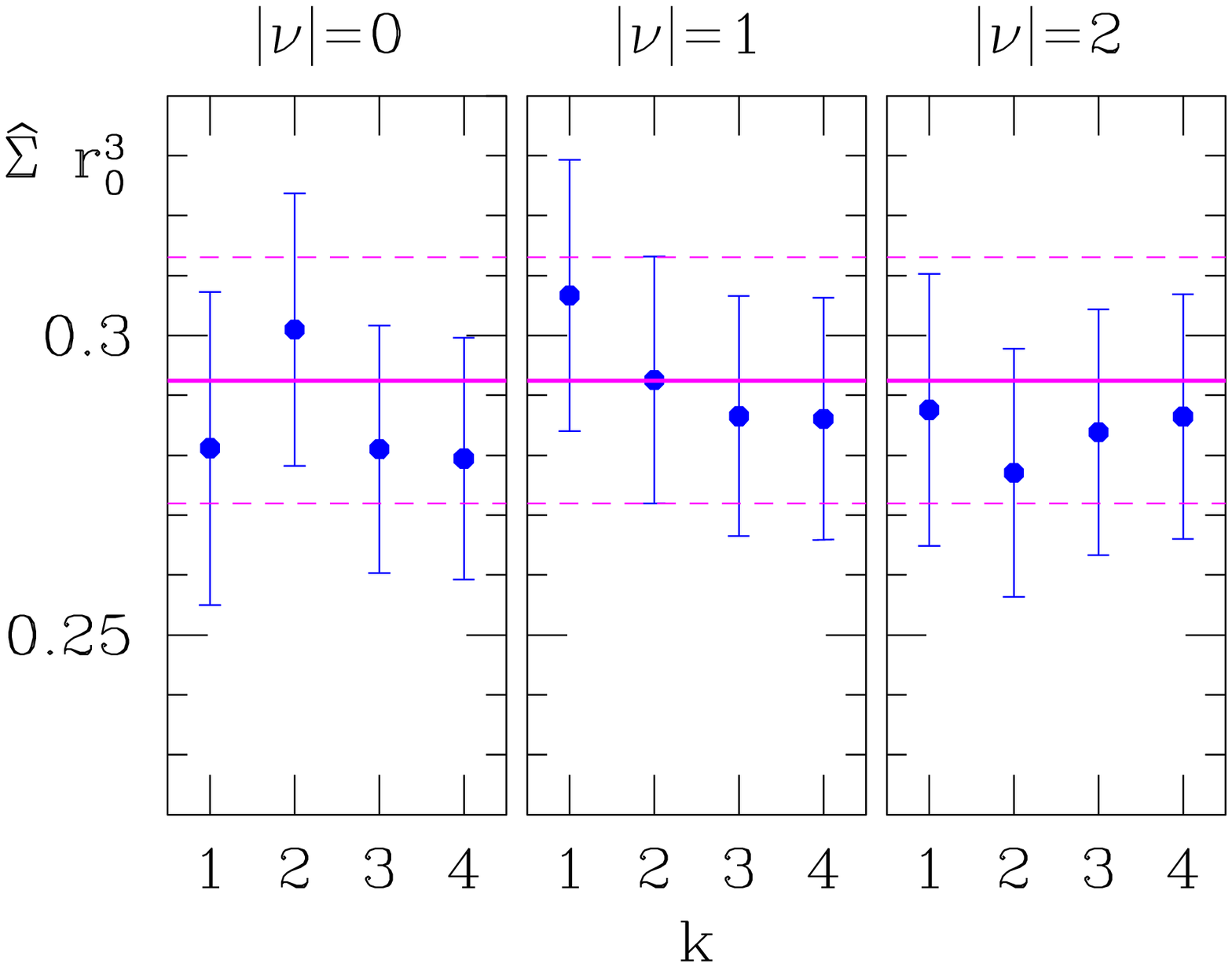}}
    \caption{\footnotesize The variation of $r_0^3\widehat\Sigma$ in
    the continuum limit, arising from choosing different eigenvalues
    and topological sectors in the determination of the bare
    condensate. The solid and dashed lines represent the result for
    $k=2,\, |\nu|=1$ which is used for our main
    result.\label{fig_fit_stab}}
  \end{minipage}
\end{figure}

Using the results for $\Sigma$ from Table~3 of \cite{rmt} (i.e. the
runs labelled $\rm B_0, B_1$ and $\rm B_2$), supplemented by our data
at $\beta=5.9256$, we plot the renormalization group invariant
condensate $\widehat\Sigma$ in units of $r_0$ versus $(a/r_0)^2$ in
Fig.~\ref{fig_zshat}. If the non-perturbative estimates for $\zshat$
are used, the results for $r_0^3\widehat\Sigma$ show a remarkably flat
behaviour, which not only indicates small residual cutoff effects, but
is also consistent with the expectation that the leading lattice
artefacts of our fermionic discretization should be of order $a^2$.

Figure~\ref{fig_zshat} also reveals that employing mean-field improved
perturbation theory for $\zshat$ produces a significant slope in
$r_0^3\widehat\Sigma$ as the continuum limit is approached. Although
this procedure apparently yields a consistent value of
$r_0^3\widehat\Sigma$ in the continuum limit, it is equally obvious
that the perturbatively renormalized result serves as a poor estimate
for the condensate at non-zero lattice spacing.

Our results for $r_0^3\widehat\Sigma$ at all values of $\beta$ and in
the continuum limit are listed in Table~\ref{tab_results}. Here we
have used $\Sigma$ as determined from $\langle\lambda_k\rangle_\nu$
for $k=2$ and $|\nu|=1$. We note that the variation in the value of
$r_0^3\widehat\Sigma$ from choosing different $\lambda_k$'s and
topological sectors is well within the statistical fluctuations after
taking the continuum limit. This is illustrated in
Fig.~\ref{fig_fit_stab}, where we plot the continuum results for all
possible choices of $\lambda_k$ and $|\nu|$. We emphasize that this
variation should not be regarded as a systematic uncertainty, since
all choices are equivalent, if random matrix theory does indeed give
an accurate description of the low-lying eigenvalues, and hence we
refrain from quoting an additional error.

\begin{table}[hb]
  \begin{center}
    \vspace{0.25cm}
    \begin{tabular}{cccc}
      \hline \\[-2.0ex] 
      $\beta$ & $r_0^3\widehat\Sigma$ & $r_0\Fk$ & $r_0m_{\rm K^*}$
      \\[0.7ex] 
      \hline \\[-2.0ex]
      $5.8458$ & $0.282(14)$ & $0.296(6)$ & $2.209(89)$ \\
      $5.9256$ & $0.285(16)$ & $0.301(7)$ & $2.403(95)$ \\
      $6.0000$ & $0.275(14)$ & $0.293(5)$ & $2.413(66)$ \\   
      $6.1366$ & $0.294(13)$ & $0.297(7)$ & $2.328(165)$ \\
      $\infty$ & ${\it 0.293(21)}$ & ${\it 0.294(9)}$ & ${\it 2.32(29)}$\\
      \hline \\[-2.0ex]
    \end{tabular}
  \end{center}
  \caption{\footnotesize Renormalization group invariant quark
    condensate, kaon decay constant and $K^*$-mass, in units of
    $r_0$.\label{tab_results}}
\end{table}

Our result in the continuum limit is thus
\be
   r_0^3\widehat\Sigma = 0.293\pm0.021
\ee
for the renormalization group invariant condensate. In the
$\msbar$-scheme at $2\,\GeV$ we obtain after division by
$\mbar_\msbar(2\,\GeV)/M=0.72076$ \cite{mbar:pap3} the value
\be
   r_0^3\Sigma_\msbar(2\,\GeV) = 0.406\pm0.029.
\ee
These are the main results of our calculation. To our knowledge, these
are the first estimates of a quantity in the continuum limit, computed
using overlap fermions. We emphasize that the quoted errors include
all uncertainties, except those due to quenching.

As is well known, the calibration of the lattice spacing is ambiguous
in the quenched approximation, and thus any conversion into physical
units is only illustrative. Here we perform such a conversion using
either the kaon decay constant or the nucleon mass to set the
scale. Ref. \cite{mbar:pap3} quotes
\be
   r_0\Fk\sqrt{2} = 0.415\pm0.009,\qquad \Fk=113\,\MeV,
\ee
in the continuum limit, while a continuum extrapolation of the
nucleon mass data of \cite{CPPAC_quen02} in units of $r_0$ yields
\be
   r_0m_{\rm N} = 2.670 \pm 0.042,\qquad m_{\rm N} = 939.6\,\MeV.
\ee
For the condensate in the $\msbar$-scheme at $2\,\GeV$ we then obtain
\be
   \Sigma_\msbar(2\,\GeV) = \left\{\begin{array}{ll}
      (285 \pm 9\,\MeV)^3, & \quad \hbox{scale set by $\Fk$} \\
      (261 \pm 8\,\MeV)^3, & \quad \hbox{scale set by $m_{\rm N}$}
                                   \end{array} \right..
\ee
These findings are consistent with previous observations that the
typical scale ambiguity for a quantity with mass dimension equal to
one is of the order of 10\%. Recent calculations of the renormalized
condensate
\cite{APE_cond,JLW_cond,RBC_cond,DeG_cond,GHR_cond,HJLW_lat01,Bern_cond,BeciLub_cond,GLMPR_cond,McNeile_cond}
yield similar values compared to our results.

%% file: sect6.tex
\section{Further scaling tests}

The leading cutoff effects of fermionic discretizations based on the
Ginsparg-Wilson relation are expected be of order~$a^2$, and indeed,
this expectation has been confirmed in our scaling study of the quark
condensate. In this section we shall extend our analysis of cutoff
effects to quantities like the pseudoscalar decay constant and the
meson mass in the vector channel.

\begin{figure}[ht]
  \begin{minipage}[t]{7cm}
    \centerline{\includegraphics*[width=7cm]{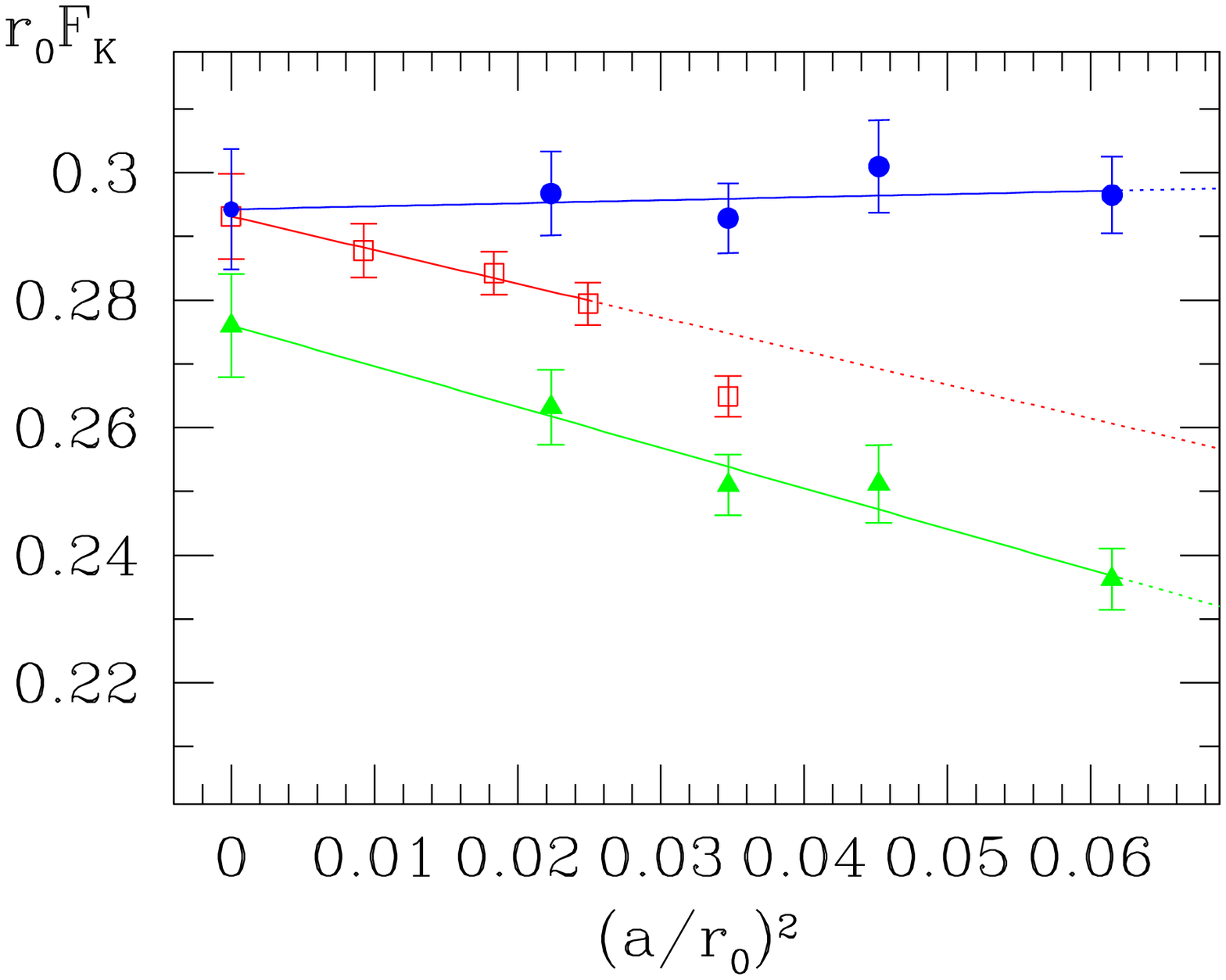}}
    \caption{\footnotesize Continuum extrapolation of $r_0 \Fk$. Full
             circles denote our results, 
             while the open squares are the data of \cite{mbar:pap3}, 
             employing O($a$) improved Wilson fermions. 
             The full triangles are our data with $\za$ from 
             mean-field improved perturbation theory. \label{FKplot}}
  \end{minipage}
  \hfill
  \begin{minipage}[t]{7cm}
    \centerline{\includegraphics*[width=7cm]{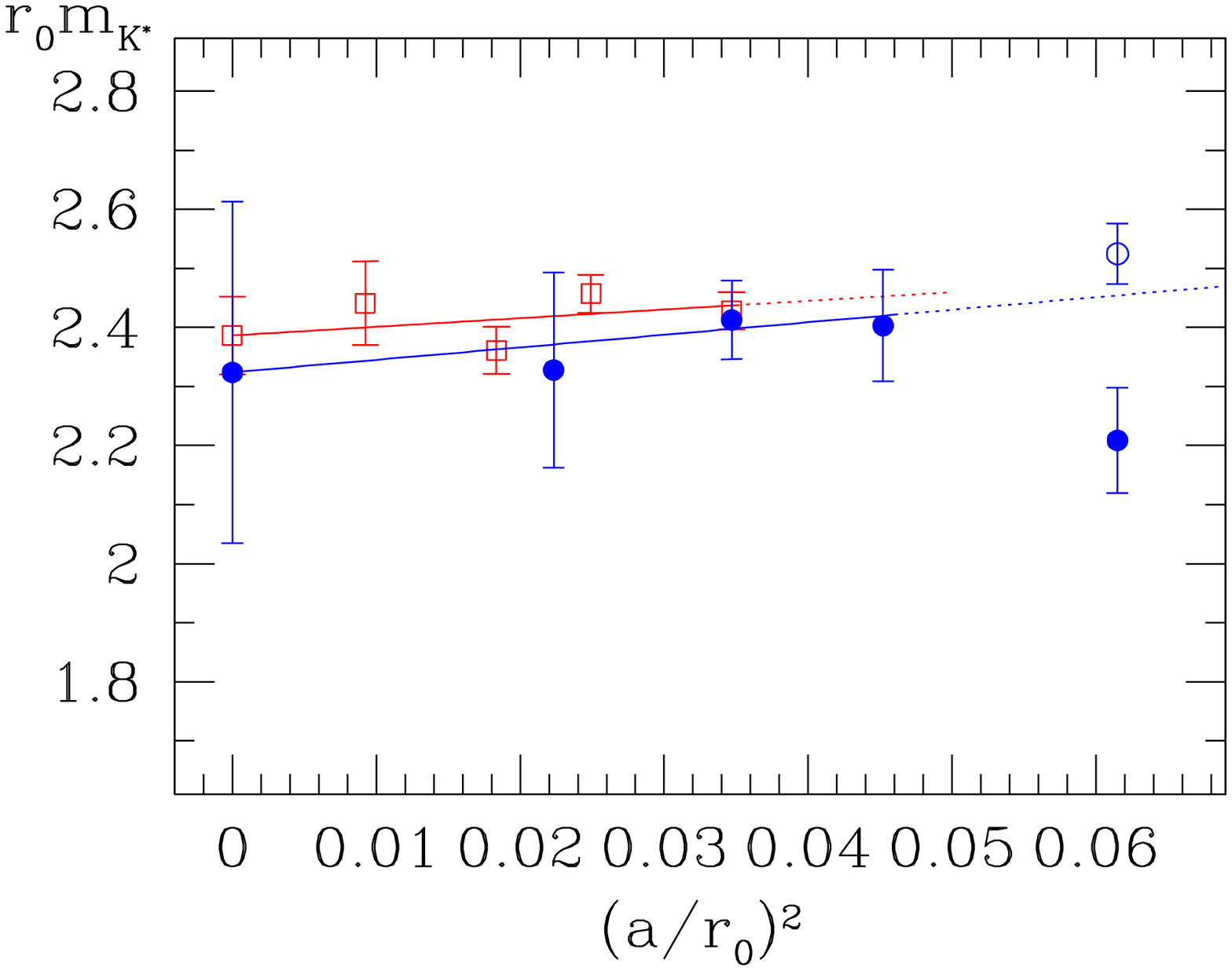}}
    \caption{\footnotesize Scaling behaviour of $r_0 m_{\rm K^*}$. The 
      meaning of the full circles and open squares is as in
      Fig.~\ref{FKplot}. The open circle results from an alternative
      fit with a fit range of $x_0/a\in[5,11]$ instead of
      $x_0/a\in[8,11]$ (full circle).
      \label{mKstarplot}} 
  \end{minipage}
\end{figure}

To this end we have assumed that $a\Fpb$ and $a\mv$ depend
linearly on $(a\mps)^2$ and performed a linear interpolation to the
point where $(r_0\mps)^2=(r_0\mk)^2=1.5736$. Thus, our aim is to
investigate the scaling behaviour of $\Fk$ and $m_{\rm K^*}$. The
renormalized kaon decay constant is obtained after multiplication with
the factor $\za$ listed in Table~\ref{tab_zs_res}. In
Table~\ref{tab_results} we have compiled the results for $r_0\Fk$ and
$r_0 m_{\rm K^*}$ at the various values of $\beta$, as well as in the
continuum limit. The corresponding continuum extrapolations are
plotted in Figures~\ref{FKplot} and~\ref{mKstarplot}.

For the kaon decay constant we observe a flat approach to the
continuum limit, consistent with a linear fit in $(a/r_0)^2$, provided
that the non-perturbative estimate for $\za$ is used.  The
perturbatively renormalized $\Fk$ is subject to larger lattice
artefacts, and the resulting continuum value is roughly consistent. In
Fig.~\ref{FKplot} we also show the continuum extrapolation of the same
quantity from ref.~\cite{mbar:pap3}, where $r_0\Fk$ was computed using
O($a$) improved Wilson fermions. In the continuum limit our data agree
remarkably well with those of ref.~\cite{mbar:pap3}, but for overlap
fermions the residual cutoff effects at lattice spacings of around
$0.1\,\fm$, i.e. at $(a/r_0)^2\approx0.035$, are apparently much
smaller.

The scaling behaviour of the $K^*$ mass is also flat, except at our
coarsest lattice spacing. A closer inspection of our fits to the
two-point function shows that the value of $a\mv$ at $\beta=5.8458$
depends strongly on the chosen fit range. Extending the fit interval
to smaller timeslices leads to a significant increase in the value of
$\r_0\mkstar$, as indicated in Fig.~\ref{mKstarplot}. Owing to the
uncertainty in the value of $\r_0\mkstar$ as a result of using
different fit intervals, we exclude the coarsest lattice from the
continuum extrapolation, despite the fact that the alternative result
is apparently consistent with a linear behaviour up to $(a/r_0)\approx
0.06$. Nevertheless we also confirm good scaling behaviour for the
vector mass; as our values for $\beta>5.8458$ are mutually consistent
with each other, as well as with the results of ref.~\cite{mbar:pap3}.

%% file: sect7.tex
\section{Conclusions}

We have presented the first comprehensive scaling study of quantities
computed using overlap fermions. A major part of our calculation was
devoted to the determination of the renormalization factor $\zshat$ of
the scalar density. Thereby we were able to present a conceptually
clean determination of the renormalized low-energy constant $\Sigma$
in the continuum limit of quenched QCD, with a total accuracy of 7\%.

Besides studying the continuum extrapolation of $r_0^3\widehat\Sigma$
we also performed scaling studies of the pseudoscalar decay constant
and the mass in the vector channel. For all three quantities computed
using overlap quarks we observed an excellent scaling behaviour,
resulting in a flat approach to the continuum limit. This is signified
by the fact that the results in Table~\ref{tab_results} at any finite
value of $\beta$ and in the continuum limit are practically the same,
at least at our level of accuracy. We note, however, that a flat
continuum behaviour is only observed for $\widehat\Sigma$ and $\Fk$,
if non-perturbative estimates of the respective renormalization
factors are employed.  Our values for $r_0\Fk$ and $r_0\mkstar$ in the
continuum limit are in very good agreement with those of
refs.~\cite{mbar:pap3,chiLF_quen}.

Owing to their good scaling properties, overlap fermions are an attractive
discretization for the computation of phenomenologically interesting 
quantities, despite the large numerical effort involved in their simulation.

\section*{Acknowledgements}
We are grateful to Leonardo Giusti, Pilar Hern\'andez, Mikko Laine,
Martin L\"uscher and Peter Weisz for interesting discussions and
for computer code developed for related projects using overlap
fermions.  We thank Miho Koma for her work on optimizing parts of our
programs. Our calculations have been performed on PC clusters at DESY
Hamburg and LRZ Munich, as well as on the IBM Regatta at FZ J\"ulich.
We thank all these institutions for support and the staff of their
computer centers for technical help.